\documentstyle[epsfig]{article}
\begin{document} \begin{center} {\Large{\bf Dynamic Transitions in Small World
Networks: Approach to Equilibrium}} \\ \vspace{1cm} Prashant M.
Gade\footnote{e-mail contact : gade@unipune.ernet.in} \\{\em Centre for
Modelling and Simulation, University of Pune, Ganeshkhind, Pune, 411 007,
India}\\ Sudeshna Sinha\footnote{e-mail contact : sudeshna@imsc.res.in} \\ {\em
The Institute of Mathematical Sciences, Taramani, Chennai 600 113, India}\\
\end{center}

\vspace{1cm}

\begin{abstract}

  We study the transition to phase synchronization in a model for the spread of
infection defined on a small world network. It was shown ( Phys. Rev. Lett.
{\bf 86} (2001) 2909 ) that the transition occurs at a finite degree of
disorder $p$, unlike equilibrium models where systems behave as random networks
even at infinitesimal $p$ in the infinite size limit. We examine this system
under variation of a parameter determining the driving rate, and show that the
transition point decreases as we drive the system more slowly. Thus it appears
that the transition moves to $p=0$ in the very slow driving limit, just as in
the equilibrium case.\\


\end{abstract}


\section{Introduction}

Dynamics of spatially extended systems has been very well studied in past two
decades. On the other hand, in the recent past, the importance of studying
networks, their structure and properties has been realized, and researchers
from fields ranging from neurodynamics and ecology to social sciences have been
extensively working in this area \cite{RMP_bar,Watts,barabasi,relev}. 
In particular, small world networks \cite{Watts} have been studied in many
different contexts. This model is defined in following way: One starts with a
structure on a lattice, for instance $k$ regular nearest neighbour connections.
Each site is now linked with $2k$ of its nearest neighbors on either side.
Then each link from a site to its nearest neighbor is rewired randomly with
probability $p$, {\it i.e.}  the site is connected to another randomly chosen
lattice site. This model is proposed to mimic real life situations in which
non-local connections exist along with predominantly local.



It has been observed in these systems, that starting from a one dimensional
chain at $p=0$, one obtains long-range order at any finite rewiring probability
with same critical exponents as in the mean-field case.  Newman and Moore
recover critical exponents for percolation on small world lattices which are
the same as for the Bethe lattice, i.e.  an infinite dimensional case
\cite{Newman}.  For XY model, Medvedyeva {\it et al} conjecture that critical
exponents are same as for the mean field case\cite{medve}. They have confirmed
it for $p \geq 0.03$ and there is good reason to believe that it is true for
any $p>0$ (The obvious difficulty is that one needs to simulate larger and
larger lattices at small $p$.)  Similar conclusions are reached for the Ising
model on a small world network.  \cite{hong}.  This strongly suggests that the
behavior for any $p\neq 0$ is the same as the behavior for $p=1$ for these
models.

Dynamical systems are nonequilibrium systems, and in general it would not be
very surprising if they have different behavior. In fact for dynamic
transitions in non-equilibrium models there is evidence of transitions at
finite $p$. For instance, the transition to self-sustained oscillations evident
in a model of infection spreading occured at finite $p$ \cite{kup}. 

Here we will try to {\em identify the conditions under which we could expect
the behavior of nonequilibrium or dynamical systems to be similar to that
observed in equilibrium models}. As a case study we use the model of infection
spreading showing finite $p$ transitions, mentioned above. First we discuss the
model in detail in Section 2.  Then in Section 3 we study the model with
respect to a parameter determining the driving rate of the system. We show how
very slow driving leads to transitions at $p \rightarrow 0$, as in equilibrium
models. We conclude in Section 4 with discussions.

\section{Model of infection spreading}

We consider the SIRS model of infection spreading on a lattice.  We take a
graph of $N$ vertices.  Each vertex has $2k$ connections.  Each site $i$ is
assigned value $\tau_i(t)$ at time $t$. The variable $\tau_i(t)$ can take
values from 0 to $\tau_0$. If $\tau_i(t)=0$, the site $i$ is considered
susceptible at time $t$.  If $\tau_I \geq \tau_i(t) \geq 1$, it is considered
infected and if $\tau_i(t)> \tau_I$ it is considered to be in the refractory
stage at time $t$.  For sites which are not susceptible, i.e. $\tau_i(t)\neq
0$, dynamics is simple:

$$\tau_i(t+1)=\tau_i(t)+1 \ \ \ \ {\rm if } \ \ \ 1\leq\tau_it(t)\leq
\tau_0-1$$ and $$\tau_i(t+1)=0 \ \ \ \ {\rm if } \ \ \ \tau_i(t)=\tau_0-1$$  

The dynamics does not depend on the neighbors if the site is not susceptible.
Neighbors come into question only while infecting the susceptible site. The
model assumes that only infected sites infect their neighbors.  Thus a site
susceptible at time $t$, will be infected at time $t+1$ with probability
proportional to the fraction of infected sites in its neighborhood. In other
words, if $\tau_i(t)=0$, $\tau_i(t+1)=1$ with the probability $p_i =
k_{inf}/k_i$ where $k_i$ are total number of neighbors of site $i$, of which
$k_{inf} $ are infected. With probability $1-p_i$, susceptible site does not
change state. 
The dynamics for the infected sites is deterministic.  The infected sites
slowly become refractory and then eventually become susceptible again.

Kuperman and Abramson simulated the above model on a small world lattice
\cite{kup}.  They observe that the fraction of infected sites at a given time
$t$ shows oscillations in time for a large value of $p$.  One can view the
system as sum of many interacting clusters and at large values of $p$, these
clusters get synchronized to each other giving collective oscillations. It was
reported in \cite{kup} that this transition to synchronization indeed occurs at
a finite value of $p$, and the transition becomes sharper in the thermodynamic
limit.

We note the following fact about phase synchronized oscillations.  If all of
them become truly synchronized, they will reach value zero at the same time and
since there are no infected sites in the lattice, infection will die down. We
want to avoid this, and hence we make a small change in the model. We add {\it
quenched disorder} or sources of infection. We choose 1\% of the total number
of sites and keep them in the infectious state forever i.e. 
$\tau_i(t)=\tau_i(0)$ for
all these sites
for all times and $\tau_i(0)=1$. This guards system against falling into fully
synchronized state where there is no further evolution\cite{footnote}.
The results in the Section below are obtained from our modified SIRS
model with quenched disorder.


\section{Results}

Specifically we study the behaviour of the infected sites with respect
to $\tau_I + \tau_R \equiv \tau_0$, which determines the rate of
driving in this model. 
We see pronounced fluctuations in the number of infected sites as a
function of time. These fluctuations are periodic with the natural
period $\tau_0$, which is the timescale for a susceptible site, if
infected, to become susceptible again.

Fig.~1 shows the time evolution of the fraction of infected elements,
for $p = 0.1$, for a particular realization after discarding a long
transient. The figure displays four cases with varying values of
$\tau_0$ (keeping the ratio $\tau_I/\tau_R$ fixed). Time has been
scaled by the natural timescale $\tau_0$ so that results from
different choices of $\tau_0$ can be easily compared. It is clear as
$\tau_0$ increases the collective oscillations get more pronounced.
These oscillations essentially indicate the presence of cycles in the
outbreak of disease.

For a small fraction of nonlocal connections $p$, these oscillations
are seen only if $\tau_0$, {\it i.e} natural timescale for the
disease, is fairly large. On the other hand, for a higher $p$, even
smaller $\tau_0$ yield collective oscillations in the number of
infected individuals at a given time. An intuitive reasoning could be
given as follows. For larger $\tau_0$ the information that a given
site is infected can propagate more, since the site stays infected for
a longer time. A similar role is played by large $p$, as the
information that a given site is infected spreads over several sites
in a very small time if one has a lot of nonlocal connections. This
sharing of information leads to collective phenomena like periodic
excitations appearing spontaneously in the system. Since higher
$\tau_0$ and $p$ play similar roles, one can expect that for higher
$\tau_0$, we will start seeing collective oscillations even at small
$p$.

As an illustrative example of the similar roles played by high $p$ and
high $\tau_0$ consider the following: the number of nonlocal
connections are certainly important in the spread of disease, as the
outbreaks can affect locations far apart geographically; but
timescales also play a role. For instance, Ebola is far more deadly
virus than HIV and kills the host much faster as it has a much shorter
incubation period. However, due to the very fact that it kills so
swiftly, Ebola outbreaks are contained very soon. The people infected
by Ebola die very quickly, and so the virus has less time to jump to a
new host and spread the disease. If no new victims come in contact
with the body fluids of infected people in their short lifetime, the
epidemic stops. On the other hand, HIV remains a problem worldwide
since victim lives longer and has longer time to infect others
\cite{ebola}.

To see this quantitatively, we study the synchronization parameter,
which is the relevant order parameter here. This is defined as
\begin{equation}
\sigma (t) = | \frac{1}{N} \Sigma^N_{j=1} \exp^{i \phi_j(t)} |
\end{equation}
where $\phi_j = 2 \pi (\tau_j -1)/\tau_0$ is a geometrical phase
corresponding to $\tau_j$. The states $\tau = 0$ are left out of the
sum \cite{kup}. As mentioned above, 1\% of the sites are quenched in
the infectious state during time evolution.  We choose initial
conditions in which 10\% of the total sites are in the infected state.
The sites which are not quenched, evolve according to the rule
mentioned above. We average over 120 configurations for $N=10^4$ and
compute the above order parameter after waiting for $2.5\times 10^4$
timesteps.

When the system is not synchronized, the phases are widely distributed
and the value of $\exp^{i \phi}$ is spread widely over the unit
circle.  This leads to small $\sigma$. On the other hand, when the
elements are synchronized, $\sigma$ is large. If all elements are
strictly synchronized, $\sigma$ will be $1$.

Fig.~2 shows the synchronization parameter $\sigma$ obtained as a time
average of $\sigma (t)$ over 1000 time steps. Subsequently we also
average over different realizations of the system. The different
curves are obtained for different values of $\tau_0$. A transition in
synchronization can be observed as $p$ runs from $0$ to $1$. This
transition occurs at values closer to $0$ as $\tau_0$ increases.
We must mention that we also carried out the same calculation for
$N=10^5$ where we averaged over 20 configurations and waited for
$9\times 10^4$ timesteps. As in case of original system, we 
observe that there is no qualitative change as we do simulations
for larger system size, except that the transition becomes 
sharper. 

The original authors postulated that transition to collective
oscillations in infected individuals could be related to the following
behavior that emerges as one increases $p$. Unlike average path
length, the clusterization decreases slowly as a function of $p$ and
there is an intermediate regime where there is a low average
clusterization $C(p)$ for a given $p$ though the distribution of
clusterization at element level $c_i(p)$ is rather broad.  They find
that this is precisely the regime when the onset of collective
oscillations occurs.  However, in this work, we studied the dynamics
of the system for different values of $\tau_0$.  This change does not
alter topology of underlying network and hence does not affect $C(p)$
or dispersion around it. But the transition is certainly affected. The
fact that the transition can be seen without changing topology of
underlying network suggests that {\em timescales also play an
  important role in this transition apart from structure of the
  network on which dynamics is taking place}.

\section{Conclusions}

We studied the transition to phase synchronization in a model for the
spread of infection defined on a small world network. It was shown in
\cite{kup} that the transition occurs at a finite degree of disorder
$p$, unlike equilibrium models where systems behave as random networks
even at infinitesimal $p$ in the infinite size limit. We examined this
system under variation of a parameter determining the driving rate,
and show that the transition point decreases as we drive the system
more slowly. Thus it appears that the transition moves to $p=0$ in the
very slow driving limit, just as one expects in the equilibrium case.

Some earlier studies may also be interpreted in this light. For
instance, it was observed that the transition point of the finite $p$
transitions to synchronization in coupled chaotic maps decreases to
$p=0$ as the chaoticity of the local map (which determines the time
scales of information loss) decreases \cite{sinha}. This can be seen
to reflect the fact that a transition at $p \rightarrow 0$ is obtained
when the rate of lyapunov exponent $1/\lambda$ tends to zero.  The
characteristic time scale for information loss in a chaotic system
varies as $1/\lambda$. So as the time scale reaches infinity, the
transition point goes to zero.  We also note Fig. 2 in our previous
paper \cite{previous}. There we have plotted the power spectra of the
collective field for small world lattices at different values of $p$.
We note that high frequency (i.e. short timescale) peaks are seen only
at large values of $p$, while small frequency (large timescale) peaks
are seen even at small $p$. Thus there is a clear interplay between
the probability of nonlocal connections $p$ and the timescales in the
system.  This suggests that in an extended parameter space one can
find dynamic transitions at infinitesimal $p$, as in the equilibrium
case, in the very slow driving parameter limit.

\begin{figure}[htb]
\label{fig1}
\begin{center}
\mbox{\epsfig{file=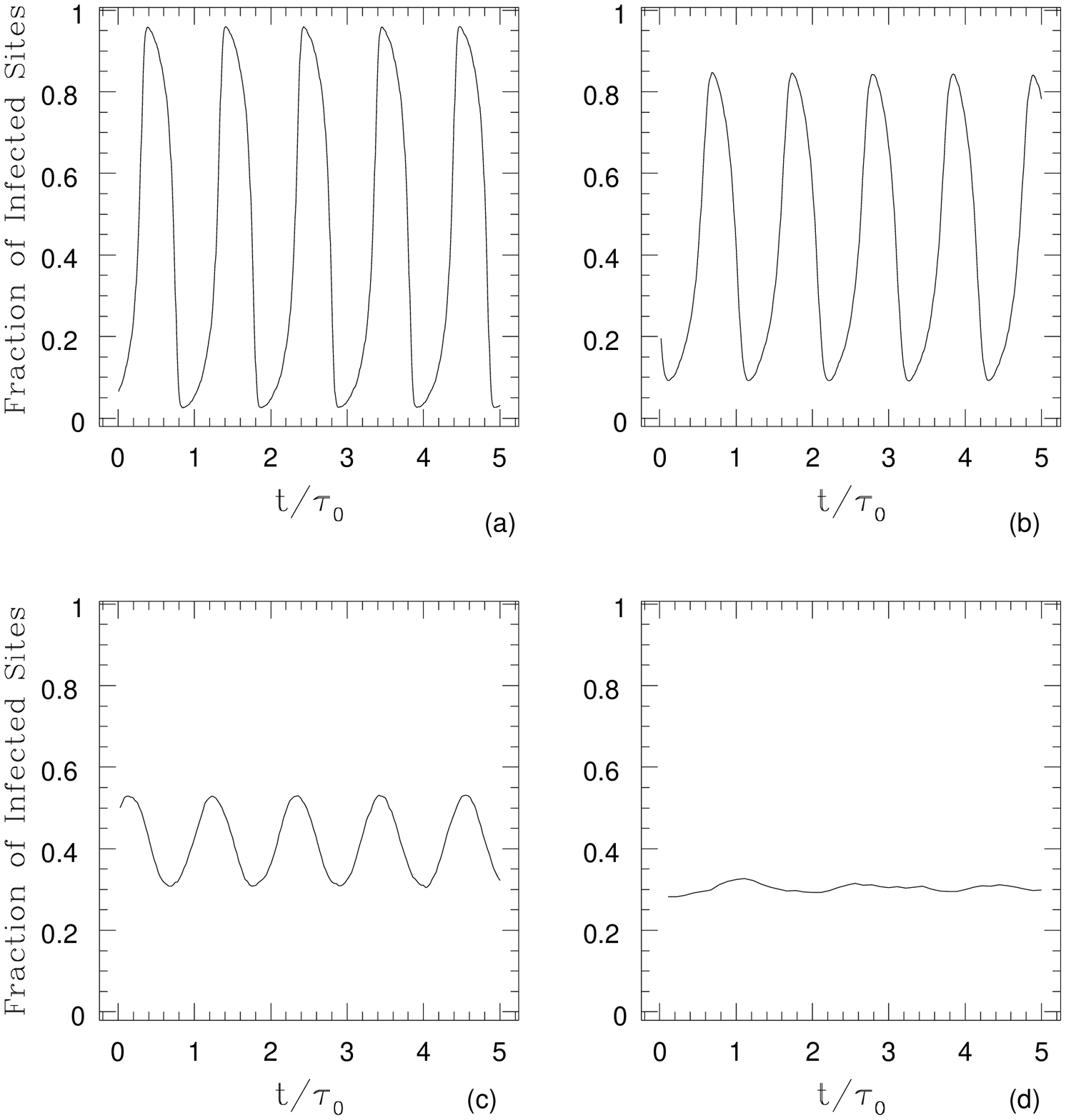,width=15truecm}}
\end{center}
\caption{Fraction of infected sites vs $t/\tau_0$ for a system of size
  $10000$, for (a) $\tau_R = 144$, $\tau_I = 64$, $\tau_0 = 208$ (b)
  $\tau_R = 72$, $\tau_I = 32$, $\tau_0 = 104$ (c) $\tau_R = 36$,
  $\tau_I = 16$, $\tau_0 = 52$ (d) $\tau_R = 9$, $\tau_I = 4$ $\tau_0
  = 13$. The value of $p$ is $0.1$.}
\end{figure}

\begin{figure}[htb]
\label{fig2}
\begin{center}
\mbox{\epsfig{file=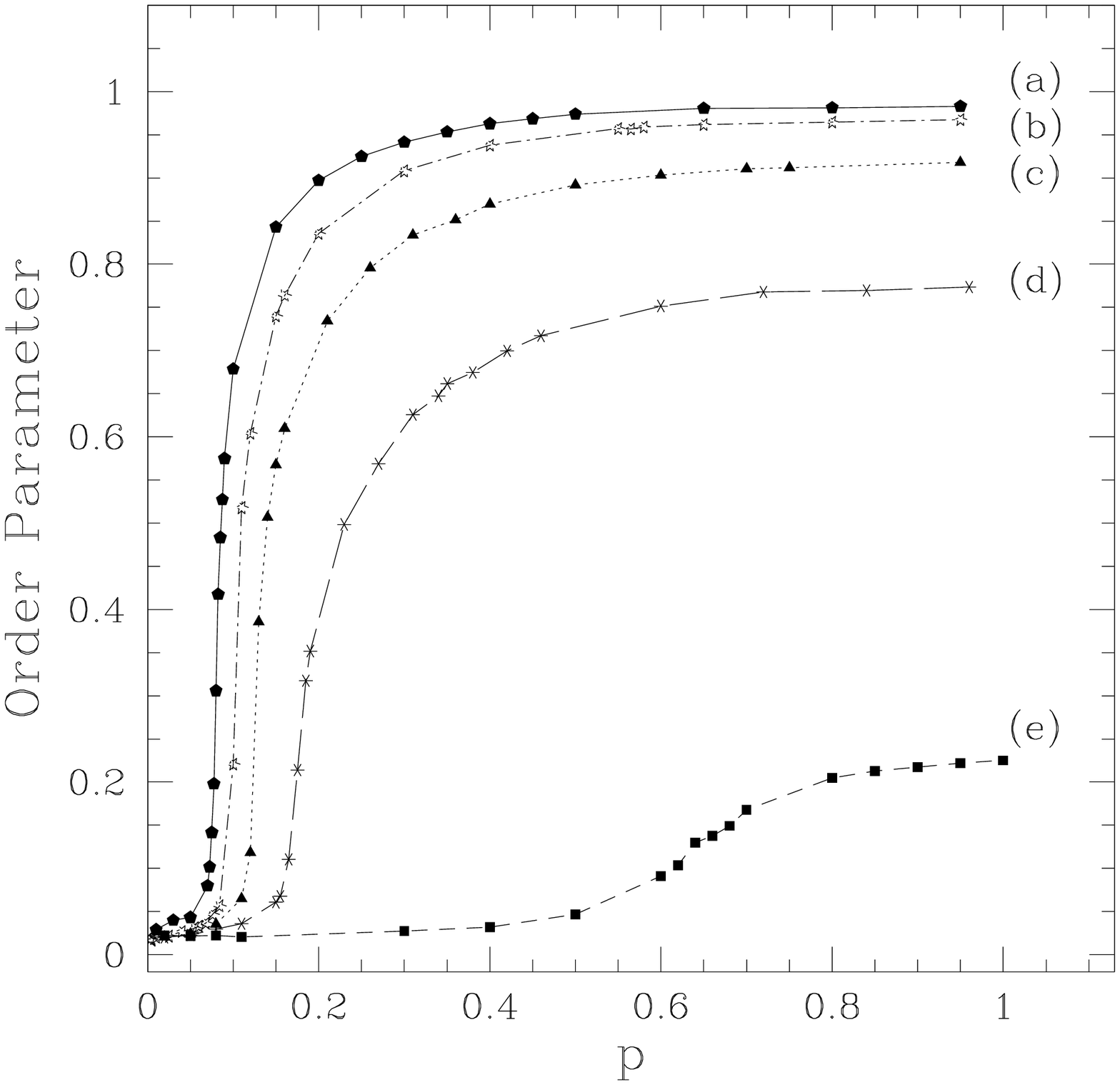,width=15truecm}}
\end{center}
\caption{Order Parameter (defined in Eqn.~1) vs $p$ for a system of
  size $10000$, for (a) $\tau_R = 144$, $\tau_I = 64$, $\tau_0 = 208$
  (b) $\tau_R = 72$, $\tau_I = 32$, $\tau_0 = 104$ (c) $\tau_R = 36$,
  $\tau_I = 16$, $\tau_0 = 52$ (d) $\tau_R = 18$, $\tau_I = 8$,
  $\tau_0 = 26$ (e) $\tau_R = 9$, $\tau_I = 4$, $\tau_0 = 13$.}
\end{figure}


\begin{thebibliography}{99}
\bibitem{footnote} Even if we do not include quenched disorder,
results for $\tau_0=9$ and $\tau_0=18$ are qualitatively similar. 
However, it is difficult to study higher $\tau_0$ since
 the possibility of synchronization and
thus system going to uninfected stage is high.
\end{thebibliography}
\end{document}